# Measurement-induced bistability in the excited states of a transmon

Jeakyung Choi[1,2,3], Hyeok Hwang[1,3], Eunseong Kim[1]*

**Abstract**

High power measurement-induced cavity response is investigated in the $|g\rangle$, $|e\rangle$, and $|f\rangle$ states of a transmon. All the states exhibit photon blockades above a certain critical value, a phenomenon that has previously been understood based on the bistability of semiclassical Duffing oscillators. The measurement-induced state transition (MIST) to high-level transmon states is expected to be one contributor to the bistability; however, the critical values measured in the $|e\rangle$ and $|f\rangle$ states are not coincident with the MIST. To understand this discrepancy, we utilize the recently developed semiclassical dynamics model of a cavity photon state. The appearance of dim and bright cavity states obtained from the model's steady-state solution leads to the photon blockades at lower critical photon numbers, and this can explain the response of the bistable region in the $|e\rangle$ and $|f\rangle$ states.

## Introduction

Quantum measurement plays an essential role in practical quantum computing applications since the outcomes of quantum computations need to be addressed through measurements of single qubits. Parity assessments in multiqubit systems are also indispensable for the realization of robust quantum error correction codes[1].

The state of a qubit is typically measured by coupling it with one or more bosonic modes and evaluating the response of the mode, which depends on the qubit state. This approach has been utilized in various quantum processing platforms including superconducting circuits and quantum dots[2,3,4,5]. The Jaynes–Cummings (JC) Hamiltonian explicates how the interaction between a qubit and cavity bosonic modes leads to the shift in the cavity resonance frequency. For instance, a substantial state-dependent detuning of the cavity frequency can be rendered by this interaction. Such changes in the cavity response allow for dispersive measurements, which minimize back-action to the qubit state and thus have been widely utilized in various quantum computing technologies[2,3,4]. When the detuning between the resonator and qubit frequency is larger than the coupling strength and the photon number is sufficiently low, the qubit–cavity system operates in the dispersive regime. By measuring dispersive shifts in the cavity frequencies, the qubit state can be identified.

[1]Korea Advanced Institute of Science and Technology, Daejeon, Republic of Korea. [2]Present address: Korea Research Institute of Standards and Science, Daejeon, Republic of Korea. [3]These authors contributed equally: Jeakyung Choi, Hyeok Hwang. *e-mail: eunseong@kaist.edu

To facilitate faster determination of qubit states, one straightforward way is to increase the measurement power to enhance the signal to noise ratio. However, an unusual cavity response appears when the photon drive of the resonator exceeds a certain critical value[6], and therefore, it is necessary to clarify the unusual response to take advantage of high-power measurements. Dips in signal amplitude caused by photon blockades have been understood using the bistability of a semiclassical Duffing oscillator[7]. Bistability has also been shown in other strongly driven transmon–resonator systems such as quantum Duffing oscillators[8,9,10], dissipative-driven JC oscillators[11,12,13,14,15,16], and Kerr nonlinear resonators[17,18,19,20]. One explanation for the origin of bistability is measurement-induced excitation between ground and excited states, attributed to the Purcell effect. On the other hand, anharmonicity of the cavity induced by interaction with the transmon, assuming first- and second-order Kerr effects, can also explain the appearance of bistability. While the key feature considered in both explanations lies in the coupling between the transmon and resonator, the two explanations take different perspectives on the interaction, namely transmon state transitions and cavity anharmonicity. Quantitative agreement between experimental results of bistability and numerical simulation has been reported, where the master equation incorporated the generalized JC Hamiltonian and photon dissipation[13]. This master equation, though, requires significant computational resources.

One effective way to test the validity of various numerical simulations is to compare the estimated critical values with those obtained in experiments since these values exhibit strong model dependence. Thus, investigation into the critical values of a transmon–cavity system is essential to understand the microscopic origin of bistability. A critical point distinguishes the dispersive and bistable regions of the cavity response. To identify the critical points, the distinctive features of photon blockades can be adopted. As the frequency changes, the stability of photon states is strongly altered, and this leads to the appearance of a photon blockade in the bistability region. Photon blockades cause distinct phase shifts and apparent amplitude dips that can be observed in the cavity response[12,13,21]. This case is clearly discriminated from conventional resonance, where amplitude peaks are accompanied by the steepest phase gradient ($d\phi/df$).

Here, to better comprehend the bistability of a transmon–cavity system, we test a different approach to find the critical points by separating the causes of bistability. We first experimentally probe the critical points where photon blockades cause a transmon–cavity system to start to exhibit unusual amplitude and phase responses by initializing the transmon in ground $|g\rangle$, first excited $|e\rangle$, and second excited $|f\rangle$ states and sweeping the power and frequency of measurement. Then to explain our experimental results we apply two models, starting with a model of qubit state excitation induced by measurement, as proposed in previous resonator-coupled two-level systems[7,21,22]. Although state transition–induced bistability has been explained for the lowest two qubit state levels, it can potentially be applied to transitions between any two other states. Recently, the measurement-induced state transition (MIST) to higher-level states has been reported[23]. In the current work, we quantitatively compare the critical points obtained in our experiment with the level crossings of the MIST beyond the rotating wave approximation (RWA) strip to investigate whether MIST level crossings could serve as an additional explanation for bistability. The second model we apply is the recently proposed semiclassical dynamics (SCD) model, developed to explain the time evolution of photon states in transmon–resonator systems and shown to demonstrate the bistable behavior of photon states[24]. We apply the SCD formula, conduct numerical simulation, and identify the critical points of our experimental bistable behavior. Taken together, the combined approaches are found to well explain the experimental critical points in a computationally efficient manner.

**Result**

**Experiment.** Our measurement setup comprises a transmon qubit capacitively coupled to an aluminum cavity at a strength of $g/2\pi$ = 55 MHz. The transmon is fabricated following the double-angle deposition technique. It has a qubit excitation frequency of $\omega_q/2\pi$ = 5.795 GHz, transmon anharmonicity of $\eta/2\pi$ = 111 MHz, cavity port coupling of $\kappa/2\pi$ = 1.3 MHz, and qubit coherence time of $T_1$ = 3.0 $\mu s$. The readout resonator has a bare frequency of $\omega_r/2\pi$ = 5.078 GHz. The coupling strength $g$ is an order of magnitude smaller than the detuning $\Delta = \omega_q - \omega_r$, which leads to dispersive coupling between the transmon and resonator when the cavity has few photons. The sample is mounted on a dilution refrigerator with a base temperature of 10 mK. A standard microwave frequency measurement setup is utilized to probe the resonance.

To initialize the transmon in the ground $|g\rangle$, first excited $|e\rangle$, or second excited $|f\rangle$ state, natural decay and 60 ns control pulses of $X_\pi^{ge}$ and $X_\pi^{gf}$ are applied, respectively. The control pulse is optimized to a $\pi$-pulse using the CRAB algorithm[25,26]. The measurement pulse, fixed at a duration of 1 μs, is a Gaussian-filtered pulse with a consistent shape that is swept by varying the amplitude of an arbitrary waveform generator (AWG) and the frequency of a signal generator (SG). Each data point represents the average of 600 measurements, and the reflection is normalized using the reflection background and converted to transmission data. The response of the transmon–cavity system is systematically examined by sweeping the measurement power and frequency, as shown in Fig. 1**a–c**. The measurement amplitude is converted to a photon number (N) at resonance by comparing with the result of two-tone spectroscopy. The purpose of this conversion is to compare the critical photon number with the value from different models.

In the low-amplitude dispersive regime, distinct resonance shifts are observed for each transmon state, as shown in Fig. 1**d**, allowing state discrimination. The Kerr effect for the ground, first excited, and second excited states is observed, and an increase in the Kerr effect is obtained as the transmon order is elevated. Here, transitions from a lower level to higher level (i.e., $|g\rangle \to |e\rangle, |e\rangle \to |f\rangle$) are generally not observed in the dispersive regime. Conversely, transitions from a higher to lower level (i.e., $|f\rangle \to |e\rangle$, $|e\rangle \to |g\rangle$) remain constant regardless of the measurement power. Therefore, in the

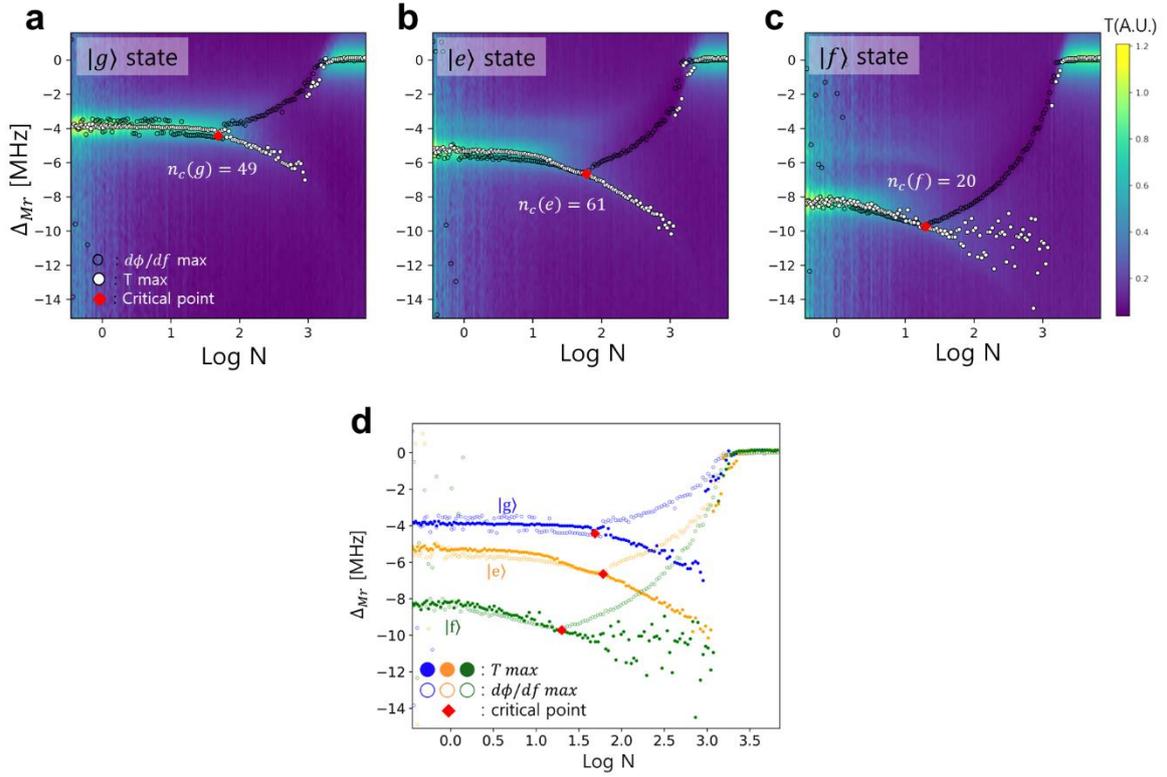

**Figure 1. Measurement result of dispersive critical points.**

**a–c** Plots of transmission data from the experimental setup for different initial states: **a** $|g\rangle$, **b** $|e\rangle$, and **c** $|f\rangle$. The y-axis is the measurement frequency detuning $\Delta_{Mr}/2\pi = (\omega_M - \omega_r)/2\pi$ and the x-axis is the logarithm of photon number *N*. As the measurement power increases, distinct dispersive, bifurcation, and punch-out regions are observed. **d** Measured peaks of the three initial states. Photon blockade signals are identified through the transmission maximum (closed circles) and the $d\phi/df$ maximum (open circles). The critical points for the $|g\rangle$, $|e\rangle$, $|f\rangle$ states are shown as the red diamonds.

dispersive regime, the state population is mainly determined by the gate fidelity of the $\pi$-pulse, $X_\pi^{ge,gf}$, and qubit decay during the measurement time.

To identify the critical point separating the photon blockade effect from the dispersive shift, we extract two sets of data points showing the maximum phase gradient ($d\phi/df$) and the maximum transmission ($T$). The critical point is then assigned as the point where the two sets start to diverge, as illustrated in Fig. 1**d**. The critical points of bistability for $|g\rangle, |e\rangle$, and $|f\rangle$ are located at $n_c(g) = 49, n_c(e) = 61, n_c(f) = 20$. To find the origin of bistability, it is necessary to examine the cavity responses with several numerical simulations.

**Models.** To find the MIST level crossings, we should calculate the eigenenergies of the states. For this, the generalized JC Hamiltonian, which considers interaction between the transmon states and cavity photon states in the absence of an external drive, is given as follows[17],

$$\frac{\widehat{H}_{GJC}}{\hbar} = \omega_r \hat{a}^\dagger \hat{a} + \omega_q \hat{b}^\dagger \hat{b} - \eta \hat{b}^\dagger \hat{b}^\dagger \hat{b} \hat{b} + g(\hat{a}^\dagger \hat{b} + \hat{a} \hat{b}^\dagger), \qquad (1)$$

where $\hat{a}$ is the cavity photon annihilation operator and $\hat{b}$ is the transmon state annihilation operator. By diagonalizing the Hamiltonian, the dressed states $|\overline{i,n}\rangle$ and their corresponding eigenenergies $E_{\overline{i,n}} = \langle \overline{i,n}|\widehat{H}_{GJC}|\overline{i,n}\rangle$ can be acquired. The overline indicates an eigenstate.

The probability of transitions between high and low transmon states becomes nontrivial in high-power regimes because of an n-dependent shift in eigenenergy. Allowed crossings between energy states can be understood by an energy ladder diagram, as shown in Fig. 2**a**. As the photon number increases, a higher transmon state energy can compared with a lower transmon state energy, and transitions beyond the RWA strip are possible by photon drive. Utilizing the eigenenergy values obtained in the experiment, the relative shift of eigenenergy from *N* photon resonator

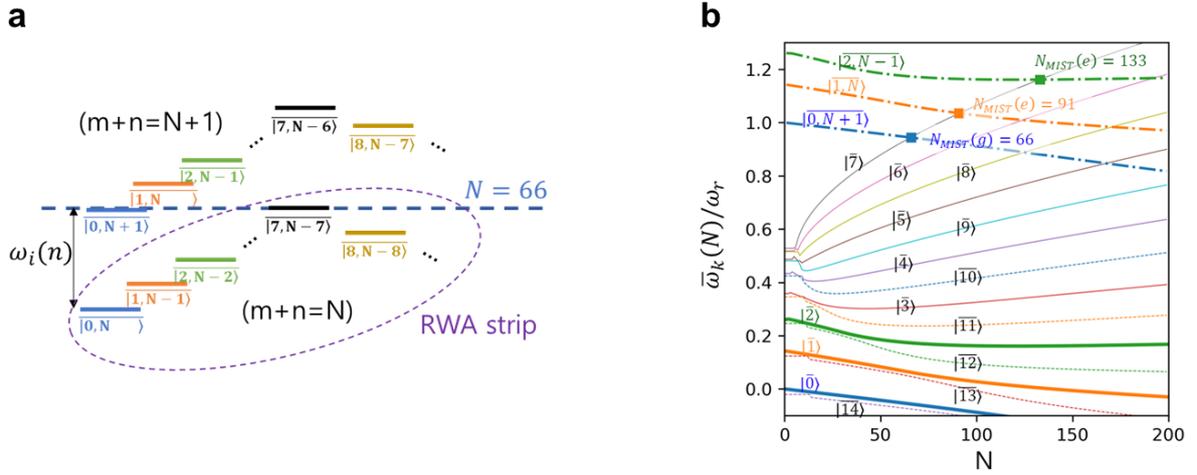

**Figure 2. Energy ladder crossing points by the MIST beyond RWA.**

**a** Energy ladder of the system showing the relative energy level of the $|k, N-k\rangle$ and $|k, N-k+1\rangle$ states. At N = 66, $|0, N+1\rangle$ and $|7, N-7\rangle$ are close in energy and the measurement drive with $\omega_i(n)$ causes a state transition between the ground and seventh excited state. **b** Fan diagram showing the relative shift $\bar{\omega}_k(N)$ of the energy level of the states divided by $\omega_r$. The blue, orange, and green solid lines respectively show the $|g\rangle$, $|e\rangle$, and $|f\rangle$ states. Third to ninth excited states are plotted with the thin solid lines, and tenth to fourteenth excited states are the dotted lines. The seventh excited state $\bar{\omega}_7(N)$ (gray solid line) shows the maximum relative energy. The dash-dotted lines plot a relative shift of the $|g\rangle$, $|e\rangle$, and $|f\rangle$ states with one additional photon ($|k, N-k+1\rangle$). The crossing points with $\bar{\omega}_7(N)$ are marked with solid squares.

($N\hbar\omega_r$) to the (k,N-k) state ($E_{\overline{k,N-k}}$) is calculated as follows[23]:

$$\bar{\omega}_k(N) = \frac{E_{\overline{k,N-k}}}{\hbar} - \omega_r N, \qquad (2)$$

where $\bar{\omega}_k(N)$ is divided by $\omega_r$ and plotted with increasing photon numbers, as shown in Fig. 2**b**. Due to the presence of cavity photon drive, $|\overline{7, N-7}\rangle$ intersects with the

$|g\rangle$ state ($|\overline{0, N+1}\rangle$) at $N_{MIST}(g) = 66$, , $|e\rangle$ state ($|\overline{1, N}\rangle$) at $N_{MIST}(e) = 91$, and $|f\rangle$ state ($|\overline{2, N-1}\rangle$) at $N_{MIST}(f) = 133$, showing crossings with high-order states. At these points, it is expected that qubit transitions occur.

The critical point for the $|g\rangle$ state ($n_c(g) = 49$) coincides with the energy crossing point in the fan diagram at $N_{MIST}(g) = 66$ within an error bound[22] of $\pm 2\sqrt{N} = \pm 16.2$, indicating that the bifurcation of the $|g\rangle$ state is due to transitions to higher-order states. However, the experimentally observed values of the bistability of the $|e\rangle$ and $|f\rangle$ states at $n_c(e) = 61$ and $n_c(f) = 20$ show discrepancy from the energy crossing points by the MIST at $N_{MIST}(e) = 91$ and $N_{MIST}(f) = 133$. This indicates that the bistability of the $|e\rangle$ and $|f\rangle$ states cannot be completely understood by the MIST.

For a specific state and photon number, the cavity resonance frequency can be estimated by the energy difference, $\hbar\omega_i(n) = E_{\overline{i,n+1}} - E_{\overline{i,n}}$. One can define effective resonance as the energy gained by the system when absorbing a single photon or the energy lost when emitting a single photon. In general, the cavity's response can be

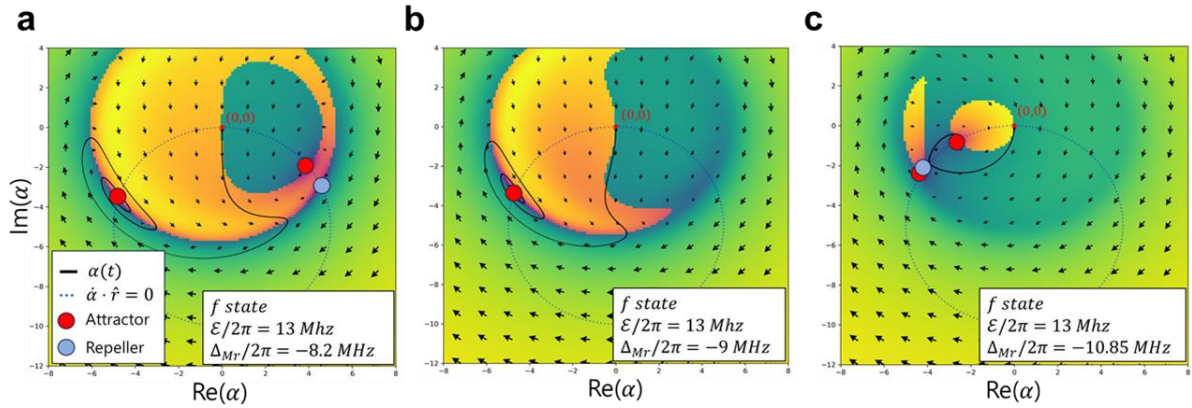

**Figure 3. Landscape of $\log|\dot{\alpha}|$ in the phase space of photon state $\alpha$.**

In each plot, $\mathcal{E}$ = 13 MHz while **a** $\Delta_{Mr}/2\pi$ = –8.2 MHz, **b** –9 MHz, and **c** –10.85 MHz. Dark and bright colors indicate low and high $|\dot{\alpha}|$. The small red dot at (0,0) is a vacuum state, the initial state of $\alpha$. The black solid line is the trajectory of $\alpha(t)$. Green and blue (orange and purple) colors show clockwise (anti-clockwise) rotation. The red circles are attractor (stable) points and the blue circles are repeller (saddle) points, where $|\dot{\alpha}| = 0$. The blue dotted circle plots $\dot{\alpha} \cdot \hat{r} = 0$. The arrows show length as $\log|\dot{\alpha}|$ and angle of $\dot{\alpha}$. **a** and **c** are high and low frequency bistable cases and **b** is a single stable case approximated to the effective resonance when ignoring other factors such as transmon state transitions and photon state dynamics. Effective resonance plays a crucial role in understanding the cavity response within the dispersive regime up to the critical point. Since the qubit frequency ($\omega_q$) is greater than the cavity frequency ($\omega_r$) in the current setup, the dispersive shift for $|g\rangle, |e\rangle, |f\rangle$ manifests a negative sign $\left(\chi_{g,e,f} = \omega_{0,1,2}(n=0) - \omega_r < 0\right)$. In the dispersive regime, this results in an effective resonance with a negative slope, referred to as the self-Kerr effect. As the photon number in the cavity increases, the slope of the effective resonance becomes positive, with the effective resonance eventually saturating to the bare frequency of the cavity

($\omega_i(n \to \infty) \to \omega_r$). These effects produce interesting local minima in the cavity response as a function of photon number. With increasing transmon state level, the dispersive shift grows with a more pronounced slope, and accordingly, the local minima move to a lower photon number. These intriguing characteristics in our arrangement allow for a rapid change of the effective resonance.

Although effective resonance is efficient in elucidating the dispersive shift and resonance changes within the dispersive regime, it does not consider transmon state transitions. For further understanding of the bistability, next we investigate the semiclassical dynamics of the cavity photons. This SCD analysis method considers the dynamical path of the cavity photons in the IQ plane that is affected by the effective resonance changes of dressed states[24]. The photon states follow semiclassical dynamics:

$$\dot{\alpha} = -i(\omega_i(|\alpha|^2) - \omega_M)\alpha - \frac{i\mathcal{E}}{2} - \frac{\kappa}{2}\alpha , \qquad (3)$$

where $\alpha$ is the coherent state of a cavity photon, $\mathcal{E}$ is the measurement drive amplitude, $\omega_M$ is the measurement drive frequency, and $\omega_i(|\alpha|^2)$ is the cavity effective frequency with transmon state *i* and photon number $n = |\alpha|^2$. The time-dependent dynamics of photon state $\dot{\alpha}$ is plotted in Fig. 3. Based on Eq. 3, $\dot{\alpha}$ shows a strong $\alpha$ dependence. The cavity photon state time evolution follows the black line in the figure, starting from a vacuum state (0,0) and saturating to a stable point.

Fig. 3**a** exhibits a high frequency bistable region. The photon state path initially approaches the dim state stable point but then bounces off near the saddle point, after which it is attracted to the bright state, showing a sudden change in amplitude and

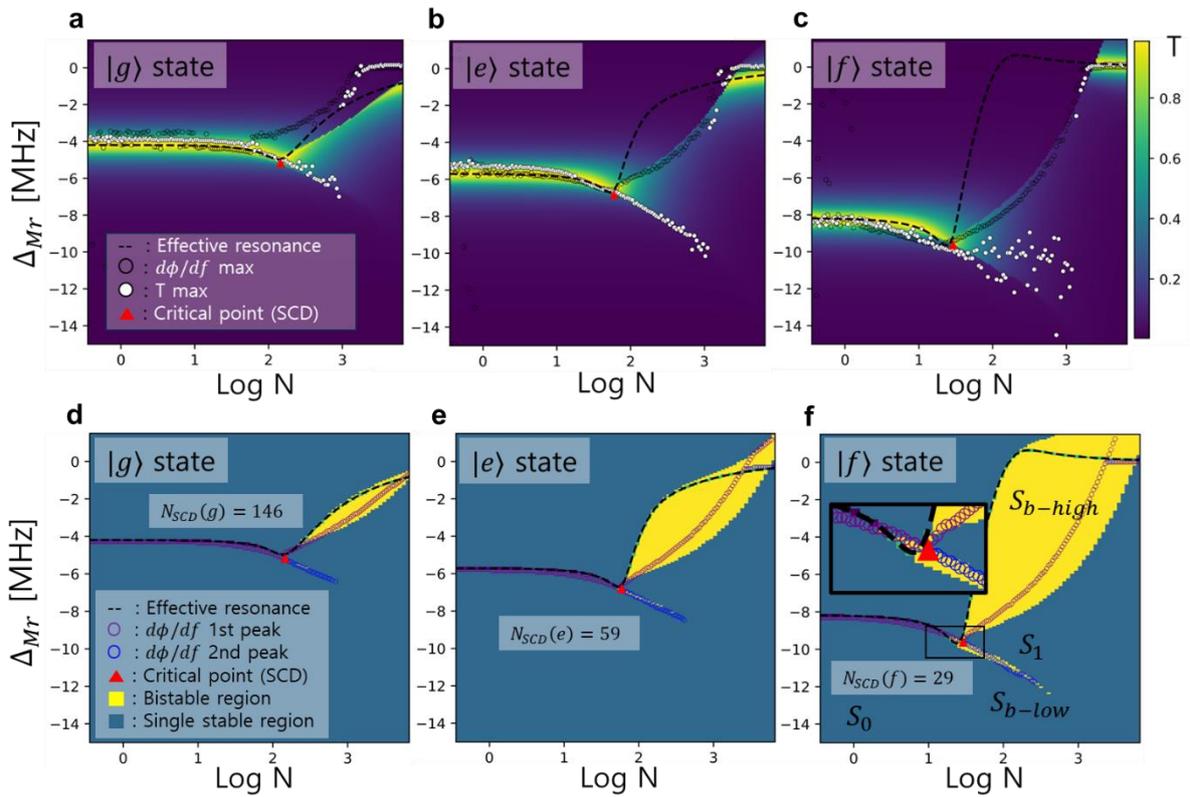

**Figure 4. SCD simulation results.**

**a–c** Comparison between the simulation results of the SCD model and measured data for $|g\rangle, |e\rangle, |f\rangle$, respectively. Measurement results are plotted as white circles ($T$ maximum) and open circles ($d\phi/df$ maximum), and the transmission amplitudes from SCD simulation are shown via the colormap. For $|e\rangle$ and $|f\rangle$, the photon blockade boundary aligns well, demonstrating that bifurcation can occur solely through semiclassical dynamics. **d–f** Plots showing the number of stable points for $|g\rangle, |e\rangle, |f\rangle$, respectively, via SCD simulation. The critical points (red triangles) can be found from the first and second phase gradient peaks from simulation (purple and blue circles). The yellow regions are bistable regions ($S_{b-low}, S_{b-high}$) having two stable points and one saddle point. The dark cyan regions ($S_0, S_1$) have a single stable point. The inset in **f** shows a magnification near the critical point.

phase. What the path drawn by the photon state shows in a series of processes is that stable points act as attractors that pull the photon state while saddle points act as repellers that push it. The transmitted signals, proportional to the photon state $\kappa\alpha$, are integrated over a fixed period of time (1 $\mu s$) during the experiment ($Te^{i\phi} \propto \int_0^{t_o} \alpha(t)dt$) and averaged over 600 times. Similarly, the amplitude and phase of the final photon state can be obtained with a fixed measurement drive and frequency in the simulation that is initiated from a vacuum state ignoring the existing photon state population. The simulation integration time is set to 1 $\mu s = 0.77/\kappa$, the same as in the experiment, and the results are plotted in Fig. 4**a–c**. With a systematic change in the measurement drive and detuning frequency, a full map of the cavity response can be obtained, which allows us to not only directly compare simulation and measurement results but also clearly identify the boundary where the stable state changes. This simulation, thus, can provide a quantitative understanding of the photon blockade boundary and the critical points of bistability (see the Supplementary Note 1 F. for more detailed discussion).

In the SCD simulation, the critical points are determined by the difference between the first and second phase gradient peaks ($d\phi/df$). The critical points from SCD are $N_{SCD}(g) = 146, N_{SCD}(e) = 59$ and $N_{SCD}(f) = 29$. Essentially the same SCD simulation results as the experiment are seen in the $|e\rangle$ and $|f\rangle$ states, while a substantial discrepancy is seen in the $|g\rangle$ state. The reason for the $|g\rangle$ state response will be discussed later.

The stable/saddle points occur where the effective resonance $\omega_i(n)$ intersects with the right side of Eq. (4). Each intersect point acts as an attractor or repeller in the photon phase space as in Fig. 3**a–c**, ($\alpha = re^{i\theta}$, $\alpha_0 = \mathcal{E}/\kappa$),

$$\omega_i(r) = \omega_M - \frac{\varepsilon}{r}\cos\theta = \omega_M \pm \frac{\kappa\alpha_0}{2r}\sqrt{1-\left(\frac{r}{\alpha_0}\right)^2}. \tag{4}$$

The number of points intersecting with $\dot{\alpha} = 0$ found in this way is plotted in Fig. 4**d–f** for $|g\rangle, |e\rangle, |f\rangle$. In all conditions, at least one intersect point occurs, which is always a stable point. In the case of three intersect points, two are stable points that act as an attractor of nonlinear dynamics while the other is a saddle point that acts as a repeller (see the Supplementary Note 1 B. for more detailed discussion).

All panels in Fig. 4**d–f** show various regions with different bistability characteristics. For instance, the number of bistable points in the simulation can be counted at a fixed measurement drive and frequency, which separates the cavity photon state into various regions. Two types of bistability with different frequency characteristics exist in the SCD method: high and low frequency regions $(S_{b-high}, S_{b-low})$. In the low-frequency region $(S_{b-low})$, both low and high amplitude stable points, often called dim and bright states, share the same sign in the phase, resulting in bistability without an apparent photon blockade. Conversely, in the high-frequency region $(S_{b-high})$, two stable points have opposite signs of phase, of which the phase difference leads to a photon blockade. Near the critical point, the dispersively shifted cavity response is further distorted by low-frequency bistability. A photon blockade appears when the measurement condition enters the high-frequency bistability region above the onset of the critical point. There is a middle region $(S_1)$ where a single stable point exists between the two bistable regions.

The first peaks of the phase gradient $(d\phi/df)$ show the resonance or the boundary between dim and bright states in $S_{b-high}$. The second peaks represent the boundary

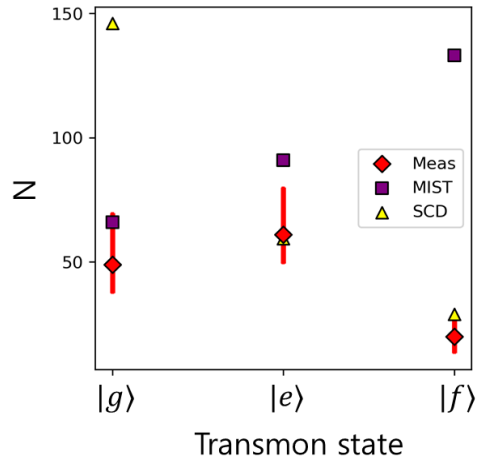

**Figure 5. Comparison of critical photon numbers in the $|g\rangle, |e\rangle$, and $|f\rangle$ states.**

The experimentally measured critical photon numbers (red diamonds) for $|g\rangle, |e\rangle, |f\rangle$ are 49, 61, and 20, respectively. Correspondingly, the MIST crossing points (purple squares) are 66, 91, and 133, and the SCD critical points (yellow triangle) are 146, 59, and 29. As the transmon level increases, the MIST crossing points decline while the critical points of SCD rise. Between the two models, the measured critical points follow the lower one in terms of photon number.

between two photon states in $S_{b-low}$. At the critical point, these types of peaks are separated. In the $|g\rangle$ state, the critical point is outside $S_{b-high}$, but in the $|e\rangle$ and $|f\rangle$ states, the critical point is in $S_{b-high}$.

When attractors in phase space exist in multiple instances, a scenario with multi-stable points can occur[12,14]. In simulation, when the system has specific conditions such as high coupling strength, there is a possibility of a small multi-stable region where two bistable regions cross. However, there is no such crossing in our system, and consequently, it is difficult to say that there are multi-stable points in the results.

The results by SCD show good agreement with those of the experimental $|e\rangle$ and $|f\rangle$ states but a clear discrepancy with the $|g\rangle$ state. To explain this discrepancy, it is helpful to note that, as previously mentioned, a photon blockade can be constructed without a qubit eigenstate transition, which is the key feature highlighted in the MIST beyond RWA approach. Fig. 5 presents a direct comparison of the critical points obtained by the experiments (red diamonds), MIST approach (purple squares) , and SCD model (yellow triangles). Measurement drive and the resultant high photon population induces nonlinear interaction between the qubit and cavity. Thus, the cavity response can be attributed to the various origins of the nonlinearity, including the MIST and SCD. In our study, the mechanism of the photon blockade changes from MIST in the $|g\rangle$ state to SCD in the $|e\rangle$ and $|f\rangle$ states. This is likely due to the fact that the appearance of low-drive bistability covers the following bistability at high measurement drive.

**Discussion**

Our results provide a quantitative explanation of the interplay between the MIST and SCD. The abrupt shift in effective resonance interrupts the photon drive, particularly when the photon state resides in a dim state. In this scenario, the photon number eventually falls short of the critical point for MIST in the $|e\rangle$ and $|f\rangle$ states. Under high measurement drive, the cavity drive frequency fails to align with the energy difference or effective resonance. Then as the system approaches punch-out, the photon number and drive closely match the effective resonance, leading to transmon ionization.

In the response of the $|g\rangle$ state, the MIST induces a critical point in the transmon state transition, causing a change in effective resonance. Consequently, the cavity response deviates from the effective resonance of the $|g\rangle$ state within the transmon state bistable region. This response exhibits a photon blockade due to the state transition, in contrast with the SCD response in the $|g\rangle$ state.

Our SCD simulation considers the dynamics within the RWA strip of the generalized JC Hamiltonian only through effective resonance and, thus, excludes transitions to other RWA strips. On the other hand, the previous MIST beyond RWA approach focused on the transition between two qubit states at high photon population while neglecting the photon state dynamics. In order to consider all these perspectives, it has been necessary to date to trace the time evolution of the generalized JC Hamiltonian while incorporating the effects of photon drive and decay. This can be achieved via numerical simulation using the Lindblad master equation with high-performance computational resources[24]. The present study, instead, suggests an effective approach to understand the dynamics of a cavity–qubit system with sufficiently low usage of computational resources.

More recently, the limitations of dispersive measurement due to the MIST within the RWA strip has been discussed[27] since photon state bistability is related to the transition of the transmon state. While that group's focus lies on the transmon state transition, the present work investigates the changes in cavity response and the bistability of the photon state. The relation between the MIST within RWA strip and the rapid change in the effective resonance is discussed in the Supplementary Note 4.

In summary, the cavity response of bistability at sufficiently high measurement drive can be understood by considering two effects of the measurement-induced transition of the transmon qubit and the resonator's effective resonance change. Both effects can induce bistability, which sets the upper boundary for the application of dispersive measurement. The interplay between both effects explains the existence of the critical photon number of dispersive measurements and describes the various intriguing cavity responses above it.

**Acknowledgement**

This work was supported by the National Research Foundation of Korea (NRF) grant funded by Ministry of Science and ICT (MSIT) (No. 2021M3E4A103887713, No. RS-2023-00282500, No. N05230124) and Korea Basic Science Institute (KBSI) National Research Facilities & Equipment Center (NFEC) grant funded by the Korean government (Ministry of Science and ICT) (No. PG2023003-07). The JPA used in this paper is supported by Irfan Siddiqi's group at UC Berkeley.

**Author contributions**

J.C. contributed the calculations and the measurements. H.H. contributed the measurements. J.C., H.H., and E.K. analyzed and discussed the results, wrote the manuscript with comments from all the authors. E.K. supervised the work.